\documentclass[prb,nofootinbib,twocolumn,superscriptaddress]{revtex4} 

\usepackage{graphicx}
\usepackage{dcolumn}
\usepackage{bm}
\usepackage{threeparttable}
\usepackage{times}
\usepackage{mathptmx}
\usepackage{lscape}
\usepackage{natbib}
\usepackage{amsmath}
\usepackage{amssymb}
\usepackage{braket}
\usepackage{comment}
\usepackage{color}

\def\degree{\kern-.2em\r{}\kern-.3em}

\begin{document}


\title{Thermodynamic Approach for Nonlinearity within Canonical Ensemble}

\author{Koretaka Yuge}
\affiliation{
Department of Materials Science and Engineering, Kyoto University, Sakyo, Kyoto 606-8501, Japan\\
}%

\begin{abstract}
{In the field of classical discrete systems, specifically substitutional alloys, this study introduces a stochastic thermodynamic approach to address nonlinearity within a canonical ensemble. This approach establishes a nonlinear relationship between a spectrum of many-body interactions and the corresponding equilibrium configuration, as determined through the canonical average. The proposed method facilitates the analysis of nonlinearity across multiple configurations via newly introduced thermodynamic functions.
These functions enable the formulation of nonlinearity in the configuration space, previously conceptualized as local, and extend it to nonlocal nonlinearity within statistical manifolds. The present findings indicate that the average nonlinearity disparity between partially ordered and other configurations is constrained by the entropy production in an ideal linear system. This system is comprehensively described by a covariance matrix of the density of states in the configuration space. 
Practically, this approach could significantly advance the analysis of nonlinearity for various classical discrete systems. 
}
\end{abstract}


\maketitle

\section{Introduction}
In the analysis of classical discrete systems characterized by $f$ structural degrees of freedom (SDFs) on a specific lattice, particularly substitutional alloys with a \textit{constant} composition, the expected structure value under a given coordination 
$\left\{ q_{1}, \cdots, q_{f} \right\}$ in thermodynamic equilibrium is determined as follows:

\begin{eqnarray}
\label{eq:can}
\Braket{ q_{p}}_{Z} = Z^{-1} \sum_{i} q_{p}^{\left( i \right)} \exp \left( -\beta U^{\left( i \right)} \right),
\end{eqnarray}
where $\Braket{\quad}_{Z}$ denotes the canonical average, $\beta$ indicates the inverse temperature, $Z=\sum_{i}\exp\left(-\beta U^{\left(i\right)}\right)$ represents the partition function, with the summation encompassing all possible configurations $i$. For instance, we can coordinate $q_{k}$ as $k$th multisite correlation function defined by the generalized Ising model,\cite{ce} which forms complete basis functions. 
Using basis functions, the potential energy $U^{\left( k \right)}$ for any configuration $k$ is expressed as follows:
\begin{eqnarray}
\label{eq:u}
U^{\left( k \right)} = \sum_{j=1}^{f} \Braket{U|q_{j}} q_{j}^{\left( k \right)},
\end{eqnarray}
where $\Braket{\quad|\quad}$ denotes the inner product in the configuration space, i.e., for any functions $a$ and $b$ of the configuration, $\Braket{a|b}=\rho^{-1}\sum_{k}a^{\left( k \right)}\cdot b^{\left( k \right)}$ with normalization constant $\rho$ and summation over all configurations. 
When we introduce two $f$-dimensional vectors of $\vec{Q}_{Z}=\left(\Braket{ q_{1}}_{Z},\cdots, \Braket{ q_{f}}_{Z}\right)$ and $\vec{U}=\left(\Braket{U|q_{1}},\cdots,\Braket{U|q_{f}}\right)$, the former and latter correspond to the configuration in thermodynamic equilibrium and many-body interatomic interactions in the inner-product form.
Subsequently, the canonical average of Eq.~\eqref{eq:can} can be interpreted as a map $\phi_{\textrm{th}}$ of
\begin{eqnarray}
\label{eq:u-q}
\phi_{\textrm{th}}: \vec{U} \mapsto \vec{Q}_{Z},
\end{eqnarray}
which is generally nonlinear. 

In the field of alloys, owing to the complex nonlinearity in $\phi_{\textrm{th}}$, a plethora of theoretical methods have been established to forecast alloy equilibrium properties. For effective exploration of the configuration space, the Metropolis algorithm was devised, followed by advanced techniques such as the multihistogram method, multicanonical ensemble, and entropic sampling.\cite{mc1,mc2,mc3,mc4} Additionally, the generalized Ising model, augmented with optimization techniques like cross-validation, genetic algorithms, and regression in machine learning, is employed to ascertain many-body interatomic interactions.\cite{cm1,cm2,cm3,cm4,cm5,cm6}
 
Although these methods, when integrated with first-principles calculations, yield precise predictions of alloy equilibrium properties, they do not inherently elucidate the nature of the canonical average for alloys as a nonlinear map. This is particularly true from the perspective of ``configurational geometry'', which is informed by the density of states in the configuration space (hereinafter referred to as ``CDOS'') on a given lattice, independent of thermodynamic variables such as temperature or energy. 

This links directly to a fundamental question in alloy configurational thermodynamics: ``How does the underlying lattice influence the overall behavior of nonlinearity in $\phi_{\textrm{th}}$ across a spectrum of equilibrium configurations and many-body interatomic interactions?'' This question remains largely unexplored. Notably, it is unclear (i) how the lattice governs nonlinearity variance among random, partially ordered, and ordered states, and (ii) how the lattice, as a spatial framework for constitutional elements, affects overall nonlocal nonlinearity.
Particularly, the importance of (i) comes from the following backgrounds: Random and candidate of ground-states (i.e. ordered states) are extreme states (the former corresponds to origin, i.e., center of gravity of CDOS, and the latter to vertices of configurational polyhedra), whose concrete configurations and their nonlinearity are therefore more easily accessible than those for partially-ordered states with diverse configurations. In this sense, averaged nonlinearity difference (particularly, averaged difference between partially ordered and other states) could be an significant measure to quantify to what extent the nonlinearity for partially ordered configurations can be deviate from that for easily-accessible other configurations, through the underlying lattice.

To address these issues, our recent theoretical investigation introduced a metric for local nonlinearity in a given configuration as a distinct vector field $\vec{H}$ in the configuration space, independent of temperature and energy.\cite{asdf,em2} We demonstrate that the magnification of the map $\phi_{\textrm{th}}$ at any given configuration, as measured from a perfectly random configuration, can be quantified by the divergence and Jacobians of this vector field.\cite{bd} Furthermore, we propose an additional metric for nonlinearity at a given configuration by expanding the concept of the vector field $\vec{H}$ to the statistical manifold.

This expansion enables the inclusion of further non-local information regarding nonlinearity, such as Kullback--Leibler (KL) divergence.\cite{ig} Based on a spatial constraint perspective, we observe a strong positive correlation between the averaged partial contribution to nonlocal nonlinearity, as measured by KL divergence across all configurations, and the geometric distance in configurational polyhedra (i.e., convex polyhedra determined from the range of correlation functions) between practical fcc binary alloys and ideally separable systems in terms of structural degrees of freedom.\cite{fcc-geom}

Although these studies have successfully introduced metrics for nonlinearity in terms of configurational geometry, they predominantly offer phenomenological insights into the overall behavior of nonlinearity, primarily due to their definition on a ``single'' configuration. Therefore, to further address the core issue of nonlinearity, a novel theoretical approach is imperative. This approach should formulate the characteristics of nonlinearity on ``multiple'' configurations based on the previously introduced nonlinearity metrics as vector field and KL divergence. 

Considering these aspects, we introduce a thermodynamic approach to nonlinearity, conceptualizing it as a stochastic system transition transformed into a system contacting with a thermal bath. This method facilitates (i) the integration of nonlinearity measures as a vector field and KL divergence via introduced thermodynamic functions, and (ii) the depiction of nonlinearity characteristics across multiple configurations through a specialized probabilistic average for thermodynamic functions and KL divergences. 

This approach unveils that the average disparity in nonlinearity between partially ordered and other configurations is limited by the entropy production in a synthetically linear system. This limitation correlates with the covariance matrix information for the density of states in the configuration space, determined exclusively by the number of multisite-figures on a given lattice. Further details are provided below.

\section{Concept and Derivation}
\subsubsection*{Basic Concepts of Nonlinearity}
Initially, we explained the fundamental concept of local nonlinearity in a configuration space as a vector field $\vec{H}$.\cite{asdf, bd}
Defining the (equilibrium) configuration as an $f$-dimensional vector $\vec{q}=\left( q_{1},\cdots, q_{f} \right)$, the vector field is expressed as follows:
\begin{eqnarray}
\label{eq:asdf}
\vec{H}\left( \vec{q} \right) = \left\{ \phi_{\textrm{th}}\left( \beta \right)\circ \left( -\beta\cdot \Gamma \right)^{-1} \right\}\cdot \vec{q} - \vec{q},
\end{eqnarray}
where $\circ$ symbolizes the composite map and $\Gamma$ is an $f\times f$ real symmetric covariance matrix of CDOS $g\left( \vec{g} \right)$.
The CDOS solely represents the probability distribution of possible configurations in the configuration space, implying $\sum_{\vec{q}}g\left( \vec{g} \right)=1$. This distribution is fundamentally independent of many-body interactions and temperature.
For simplicity, we assume the origin of the configuration space as a perfectly random configuration $\vec{q}_{0}$, defined as the center of gravity of the CDOS.
 
In terms of configurational geometry, we have demonstrated that $\phi_{\textrm{th}}$ exhibits a globally linear map when the CDOS assumes a multidimensional Gaussian form.\cite{ig} 
This linearity can be confirmed by directly estimating the canonical average with the Gaussian CDOS,\cite{em2} and can also be understood by employing series expansion of the canonical average in terms of moments of the CDOS: The latter clearly states that to achieve the global linearity for practical system, relationship between $t$-th $\left( t \ge 3 \right)$ order moments and the 2nd order moments for practical CDOS should be the same as the Gaussian.\cite{ig} Therefore, we can naturally start from considering the nonlinearity in $\phi_{\textrm{th}}$ as any deviation in practical CDOS from the corresponding Gaussian.  
This study also reveals that local nonlinearity in $\phi_{\textrm{th}}$ at configuration $\vec{q}$ can be decomposed into linear and nonlinear contributions, with the former characterized by the invertible map of $\left( -\beta\cdot\Gamma \right)$. Considering these points, we highlight key aspects of the vector field $\vec{H}$: (i) when $\phi_{\textrm{th}}$ is locally linear at configuration $\vec{q}$, $\vec{H}\left( \vec{q} \right)$ assumes a zero-vector, and (ii) the image of the composite map $\phi_{\textrm{th}}\left( \beta \right)\circ \left( -\beta\cdot \Gamma \right)^{-1}$ is independent of temperature and many-body interactions. Thus, the local nonlinearity of $\vec{H}\left( \vec{q} \right)$ is unaffected by temperature and energy, meaning $\vec{H}$ can be \textit{a priori} determined based solely on CDOS, $g\left( \vec{q} \right)$. 

We then extend the concept of $\vec{H}$ in the configuration space to the statistical manifold, thereby capturing additional nonlocal nonlinearity at the specific configuration $\vec{q}$. The corresponding nonlocal nonlinearity at configuration $\vec{q}_{A}$, $D_{\textrm{NOL}}^{A}$, is defined by the KL divergence of:\cite{ig} 

\begin{eqnarray}
D_{\textrm{NOL}}^{A} = D_{\textrm{KL}}\left( c_{A} : c_{\textrm{G}A} \right) = \sum_{\vec{q}} c_{A}\left( \vec{q} \right) \ln \frac{c_{A}\left( \vec{q} \right)}{c_{\textrm{G}A}\left( \vec{q} \right) }.
\end{eqnarray}
$c_{A}$ and $c_{\textrm{G}A}$ are probability distributions for configuration $\vec{q}_{A}$, given by
\begin{eqnarray}
\label{eq:cdoss}
c_{A} \left( \vec{q} \right)&=& \dfrac{g\left( \vec{q} \right)\exp\left[ -\beta\left( \vec{q}\cdot \vec{V}_{A} \right) \right]}{ \sum_{\vec{q'}}g\left( \vec{q'} \right)\exp\left[ -\beta\left( \vec{q'}\cdot \vec{V}_{A} \right) \right] } \nonumber \\
c_{\textrm{G}A} \left( \vec{q} \right)&=& \dfrac{g_{\textrm{G}}\left( \vec{q} \right)\exp\left[ -\beta\left( \vec{q}\cdot \vec{V}_{A} \right) \right]}{ \sum_{\vec{q'}}g_{\textrm{G}}\left( \vec{q'} \right)\exp\left[ -\beta\left( \vec{q'}\cdot \vec{V}_{A} \right) \right] },
\end{eqnarray}
where $g\left( \vec{q} \right)$ corresponds to the CDOS of practical system with covariance matrix $\Gamma$, $g_{\textrm{G}}\left( \vec{q} \right)$ corresponds to the CDOS of ``linear system'', given by multidimensional Gaussian with the same $\Gamma$, and 
\begin{eqnarray}
V_{A} = \left( -\beta\cdot\Gamma \right)^{-1}\cdot \vec{q}_{A}.
\end{eqnarray}

\begin{figure}[h]
\begin{center}
\includegraphics[width=0.7\linewidth]{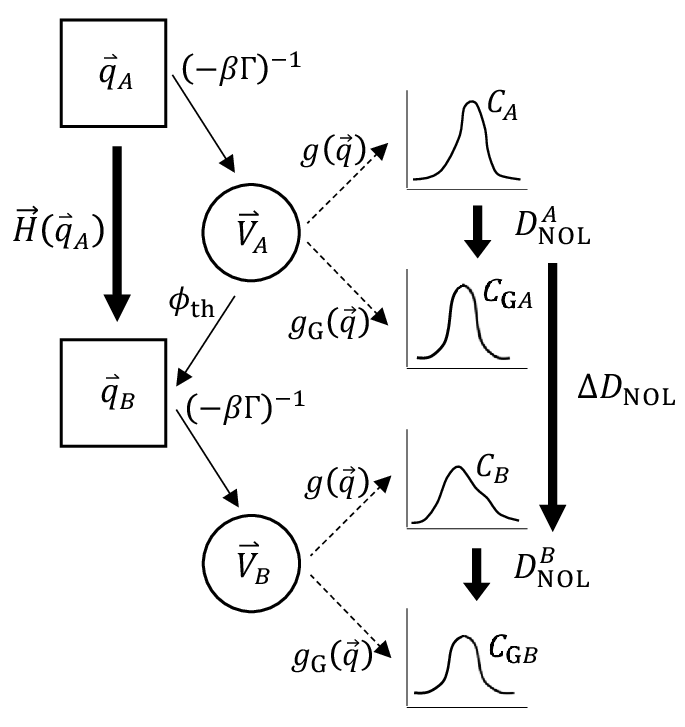}
\caption{Relationship between nonlinearity on configuration space as a vector $\vec{H}$ at configuration $\vec{q}_{A}$ and nonlinearity on statistical manifold as KL divergence $D_{\textrm{NOL}}$.  Thin arrows denote taking map (e.g., $\vec{V}_{A} = \left( -\beta\Gamma \right)^{-1}\cdot \vec{q}_{A}$ and $\vec{q}_{B} = \phi_{\textrm{th}}\cdot \vec{V}_{A}$), dashed arrows denote the operation of canonical average with $\vec{V}_{J}$ and CDOS ($g$ or $g_{\textrm{G}}$) (i.e., corresponding to Eq.~\eqref{eq:cdoss}), and bold arrows represent differences on configuration space (as vector) or on statistical manifold (as Kullback-Leibler divergence), i.e., 
$\vec{H}\left( \vec{q}_{A} \right) = \vec{q}_{B} - \vec{q}_{A}$ and $\Delta D_{\textrm{NOL}}=D_{\textrm{NOL}}^{B}-D_{\textrm{NOL}}^{A}$.  }
\label{fig:nol-world}
\end{center}
\end{figure}

For example of equiatomic binary system on fcc lattice with pair correlations, corresponding linear system has the Gaussian CDOS with diagonal covariance matrix $\Gamma$: In this case, U-Q correspondence in Eq.~\eqref{eq:u-q} for the linear system is explicitly given by 
\begin{eqnarray}
\Braket{q_{k}}_{Z} = -\beta\Gamma_{kk} \Braket{U|q_{k}},
\end{eqnarray}
where $\Gamma_{kk}$ denotes $k$-th diagonal element of $\Gamma$. For the practical fcc binary system, this linearity generally does not hold, i.e., $\left\{\Braket{q_{k}}_{Z}\right\}$ cannot be obtained through constant matrix operation on $\left\{\Braket{U|q_{k}}\right\}$. 
Hereafter, we employ the subscript $\textrm{G}$ as a function of the linear system, as defined for $c_{\textrm{G}A}$ and $g_{\textrm{G}}$.  
Note that in order to practically measure the nonlinearity for classical discrete systems, we here consider that the linear system has CDOS of discrete Gaussian and resultant discrete canonical distributions: We have confirmed that this linear system can practically hold local linearity in terms of the zero-vector field inside the configurational polyhedra, when support of the practical CDOS is included in that of discrete Gaussian CDOS for linear system.\cite{cls}   
We can also see that $D_{\textrm{NOL}}^{A}$ is independent of the temperature and many-body interactions, which is a common characteristic with $\vec{H}$.
The relationships in nonlinearity between vector of $\vec{H}$ on configuration space ($\vec{q}_{B}=\vec{q}_{A} + \vec{H}\left( \vec{q}_{A} \right)$) and  KL divergence of $D_{\textrm{NOL}}$ on statistical manifold is depicted in Figure~\ref{fig:nol-world}. 
$\Delta D_{\textrm{NOL}} = D_{\textrm{NOL}}^{B} - D_{\textrm{NOL}}^{A}$ denotes the difference in nonlocal nonlinearity between configurations $\vec{q}_{B}$ and $\vec{q}_{A}$. 


\subsubsection*{Preparations for Thermodynamic Treatment of Nonlinearity}

As discussed earlier, there are two main challenges: (i) both $\vec{H}$ and $D_{\textrm{NOL}}$ are defined within a single configuration, thus not directly formulating the behavior of nonlinearity across multiple configurations, and (ii) given that $\vec{H}$ and $D_{\textrm{NOL}}$ are delineated in distinct spaces, their integrated treatment remains ambiguous. 
To address these issues, our study proposes a thermodynamic approach to nonlinearity, comprising the following steps:

\begin{figure}
\begin{center}
\includegraphics[width=0.72\linewidth]{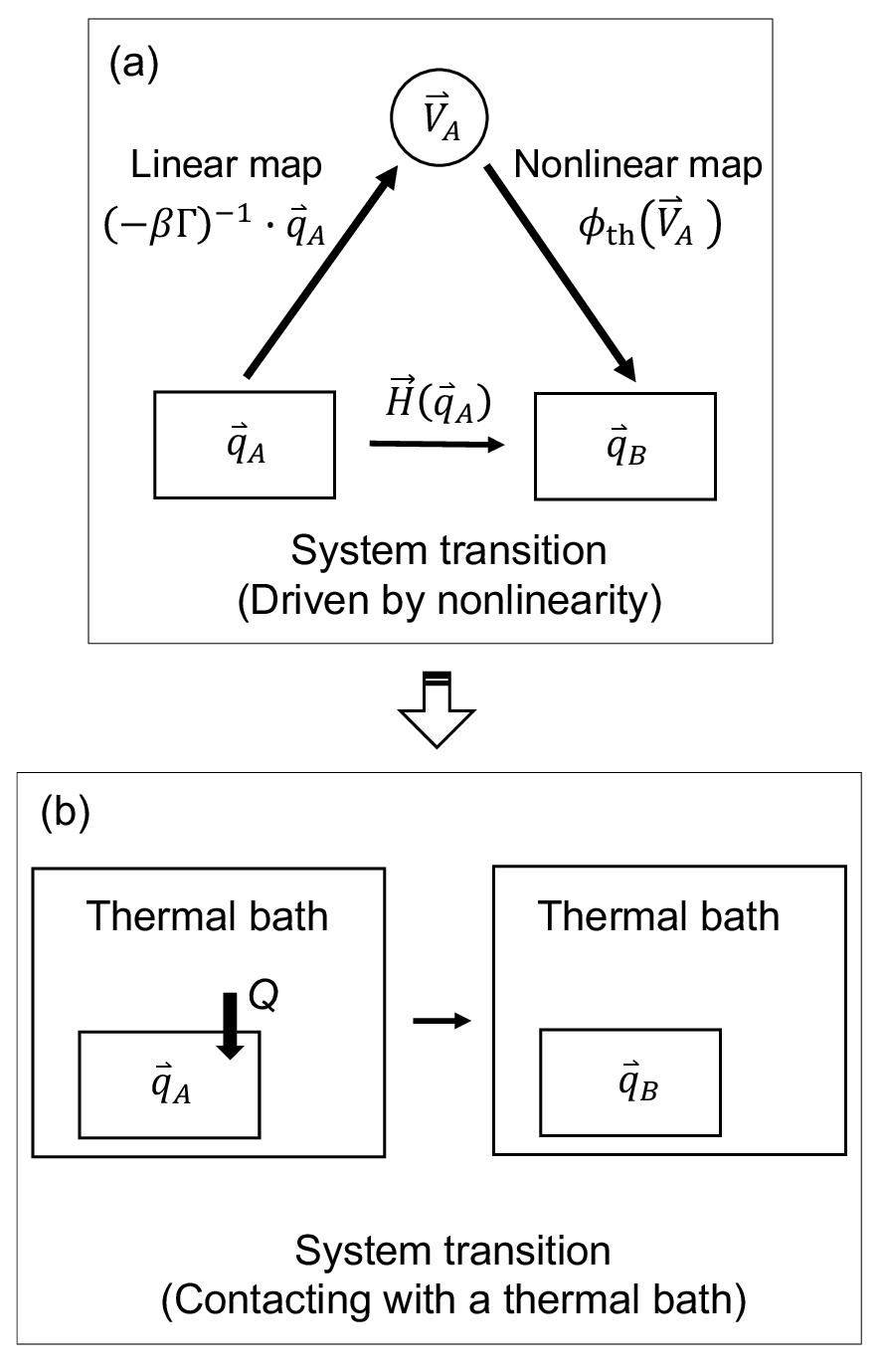}
\caption{Schematic illustration of (a): System transition from $\vec{q}_{A}$ to $\vec{q}_{B}$ on configuration space driven by nonlinearity, transformed into (b): System transition contacting with a  thermal bath. To achieve the transformation, extension of the deterministic system transition to stochastic one is required, as described in Eqs.~\eqref{eq:gp}-\eqref{eq:dt-r}.}
\label{fig:cp-trans}
\end{center}
\end{figure}

First, we consider local nonlinearity $\vec{H}$ as the time evolution of the dynamical system: 
\begin{eqnarray}
\label{eq:dynamical}
\vec{q}_{t+1} = \vec{q}_{t} + \vec{H}\left( \vec{q}_{t} \right).
\end{eqnarray}
This corresponds to interpreting the local nonlinearity $\vec{H}$ at $\vec{q}_{t}$ as the transition $\vec{q}_{t}\to \vec{q}_{t+1}$ driven by the nonlinearity from time $t$ to $t+1$ in the configuration space. 

Subsequently, the dynamical system is extended into a stochastic transition by expanding the system configuration $\vec{q}$ at time $t$ to a probability distribution $P\left( \vec{q} \right)$. The transition probability from the initial configuration $\vec{q}_{A}$ at time $t$ to the final configuration $\vec{q}_{B}$ at time $t+1$ is represented by
\begin{eqnarray}
\label{eq:gp}
R\left( \vec{q}_{B}|\vec{q}_{A} \right)= \cfrac{g\left( \vec{q}_{B} \right) \exp\left[ -\beta \left( \vec{q}_{B}\cdot \vec{V}_{A} \right) \right]} {\sum_{\vec{q}} g\left( \vec{q} \right) \exp \left[ -\beta \left( \vec{q}\cdot \vec{V}_{A} \right) \right] },
\end{eqnarray}
which is time independent. Based on this definition, we can determine the relationship $R\left( \vec{q}_{B}|\vec{q}_{A} \right) = c_{A}\left( \vec{q}_{B} \right)$. Hereafter, the denominator of Eq.~\eqref{eq:gp} as $Z_{A}$:
Therefore, $R\left( \vec{q}_{B}\middle| \vec{q}_{A} \right)$ can be interpreted as probability of finding configuration $\vec{q}_{B}$  under canonical distribution with many-body interaction $\vec{V}_{A}$ obtained through linear map onto $\vec{q}_{A}$, i.e., $\vec{V}_{A}=\left( -\beta\cdot\Gamma \right)^{-1}\cdot \vec{q}_{A}$.
Accordingly, the definition of the transition probability in Eq.~\eqref{eq:gp} recovers the original dynamical system by considering the ensemble average of the final configurations,
\begin{eqnarray}
\label{eq:dt-r}
\sum_{\vec{q}'_{B}}R\left( \vec{q}'_{B} \middle| \vec{q}_{A} \right)\cdot \vec{q}'_{B} = \vec{q}_{A} + \vec{H}\left( \vec{q}_{A} \right),
\end{eqnarray}
indicating that we can naturally extend the dynamical system driven by nonlinearity in Eq.~\eqref{eq:dynamical} into a stochastic system transition using Eq.~\eqref{eq:gp}.

Finally, we reconceptualized the stochastic transition driven by nonlinearity in the configuration space as a system transition interacting with a thermal bath at an inverse temperature $\beta'$. Given that the characteristics of nonlinearity, represented by the vector field $\vec{H}$ and KL divergence $D_{\textrm{NOL}}$, are independent of $\beta$, we equate $\beta=\beta'$. 

With already progressed from a deterministic to a stochastic transition, the state of the system at time $t$ is characterized by a probability distribution. In this context, and referring to stochastic thermodynamics,\cite{ft} we focus on the transition from the initial configuration $\vec{q}_{A}$ at time $t$ to the final configuration $\vec{q}_{B}$ at time $t+1$, which permits the definition of thermodynamic and related functions post-transformation. For instance, the variation in system entropy $\Delta S_{\textrm{sys}}$ is determined as follows:

\begin{eqnarray}
&&\Delta S_{\textrm{sys}} = \ln\frac{P\left( \vec{q}_{A} \right)}{ P'\left( \vec{q}_{B} \right)}, 
\end{eqnarray}
where $P$ and $P'$ denote the probabilities of the system state at times $t$ and $t+1$, respectively. 
Similarly, the bath entropy change $\Delta S_{\textrm{b}}$, $f_{J}$ at $\vec{q}_{J}$ and the entropy production $\sigma$ is defined as  
\begin{eqnarray}
\Delta S_{\textrm{b}} &=& \ln\frac{R\left( \vec{q}_{B} | \vec{q}_{A}\right)}{R\left( \vec{q}_{A} | \vec{q}_{B} \right) } \nonumber \\
f_{J} &=& -\beta^{-1}\ln Z_{J} \nonumber \\
\sigma &=& \Delta S_{\textrm{sys}} + \Delta S_{\textrm{b}}. 
\end{eqnarray}
Hereafter, for any configurational function $W\left( \vec{q} \right)$, we considered $\Delta W$ as the difference in $W$  between the initial  
$\vec{q}_{A}$ and final configurations $\vec{q}_{B}$, namely, 
$\Delta W = W\left( \vec{q}_{B} \right) - W\left( \vec{q}_{A} \right)$.
Noted that $f_{J}$ does not correspond to the free energy \textit{after} the transformation illustrated in Fig.~\ref{fig:cp-trans}. 
These functions yield the corresponding forward and backward transition probabilities of 
\begin{eqnarray}
P_{+}\left( \vec{q}_{A}, \vec{q}_{B} \right) &=& R\left( \vec{q}_{B}| \vec{q}_{A} \right) P\left( \vec{q}_{A} \right) \nonumber \\
P_{-}\left( \vec{q}_{A}, \vec{q}_{B} \right) &=& R\left( \vec{q}_{A}| \vec{q}_{B} \right) P'\left( \vec{q}_{B} \right),
\end{eqnarray}
where the second law of thermodynamics, $\Braket{\sigma}_{P{+}}\ge 0$, is inherently satisfied because of the positive semidefinite character of the KL divergence.
To connect the thermodynamic functions with the nonlinearity, we introduce the concept of ``stochastic eigen nonlinearity'' at any given  configuration $\vec{q}_{I}$, defined as 
\begin{eqnarray}
d_{I} = \ln R\left(\vec{q}_{I}|\vec{q}_{0}\right),
\end{eqnarray}
which significantly influences the measurement of the nonlinearity in the context of thermodynamic treatments, as discussed later. 
Thereafter, the corresponding variations in $d$ through the transition $\vec{q}_{A}\to\vec{q}_{B}$ can be described as follows: 
\begin{eqnarray}
\Delta d = d_{B} - d_{A} = \ln\frac{R\left(\vec{q}_{B}|\vec{q}_{0}\right) }{R\left(\vec{q}_{A}|\vec{q}_{0}\right)}. 
\end{eqnarray}
We defined the special forward and backward transition probabilities, where the state probability distribution accounts for CDOS itself: 
\begin{eqnarray}
\label{eq:p0}
P_{+}^{0} &=& R\left( \vec{q}_{B}| \vec{q}_{A} \right) g\left( \vec{q}_{A} \right) \nonumber \\
P_{-}^{0} &=& R\left( \vec{q}_{A}| \vec{q}_{B} \right) g\left( \vec{q}_{B} \right),
\end{eqnarray}
and the corresponding entropy production through transition from $\vec{q}_{A}$ to $\vec{q}_{B}$ is introduced as follows: 
\begin{eqnarray}
\sigma^{0} \left( \vec{q}_{A}, \vec{q}_{B} \right) = \ln\frac{P_{+}^{0}}{P_{-}^{0} } = d_{\textrm{KL}}\left( P_{+}^{0}, P_{-}^{0} \right) ,
\end{eqnarray}
where $d_{\textrm{KL}}$ denotes the stochastic relative entropy.

Note that through the thermodynamic transformation, Hamiltonian of the system is not known \textit{a priori}. From Eq.~\eqref{eq:gp}, while $\vec{V_{A}}$ appears acting as interaction, a set of $\left\{ \vec{q}\cdot\vec{V}_{A} \right\}$ or $\left\{ \vec{q}_{B}\cdot\vec{V}_{A} \right\}$ does not correspond to the Hamiltonian: They are introduced as ``artificial interaction'' in order to merely measure the nonlinearity in $\phi_{\textrm{th}}$. Therefore, in the context of thermodynamics, Hamiltonian of the system would be determined from steady state (through the transition probability of Eq.~\eqref{eq:gp}) satisfying the detailed balance condition. 

Based on these preparations, when we synthesize the notions of nonlinearity $\vec{H}$ in the configuration space and $D_{\textrm{NOL}}$ in the statistical manifold, the introduced thermodynamic and related functions are appraised in comparison to those of a linear system. This comparison is logically inferred from the relationship between $\vec{H}$ and $\Delta D_{\textrm{NOL}}$ as illustrated in Fig.~\ref{fig:nol-world}.
Henceforth, the tilde symbol denotes a function of the practical system measured against the linear system, for example, $\tilde{W}_{I}=W_{I} - W_{\textrm{G}I}$ and $\Delta \tilde{W} = \tilde{W}_{B} - \tilde{W}_{A}$.

\subsubsection*{Unified Treatment of Nonlinearity through Thermodynamics}
In this work, we first establish a connection between nonlocal nonlinearity, as represented by KL divergence on the statistical manifold, and local nonlinearity, depicted as a vector field on the configuration space, through the thermodynamic quantities introduced.

To facilitate this, we introduce a novel average for a function of the initial and final configurations of $M\left( \vec{q}_{A}, \vec{q}_{B} \right)$, defined as

\begin{eqnarray}
\label{eq:VA}
\Braket{M}_{{AB}} &=&  \sum_{\vec{q}'_{B}} R\left( \vec{q}'_{B}|\vec{q}_{A} \right) \cdot M\left( \vec{q}_{A}, \vec{q}'_{B} \right) + \sum_{\vec{q}'_{A}} R\left( \vec{q}' _{A} | \vec{q}_{B}\right) \cdot M\left( \vec{q}'_{A},\vec{q}_{B} \right) \nonumber \\
 &=& \Braket{M}_{{A}} + \Braket{M}_{{B}}.
\end{eqnarray}
Based on this definition, $\Braket{M}_{A}$ ($\Braket{M}_{B}$) signifies the canonical average of $M$ over the final (initial) configuration under many-body interaction at the initial (final) configuration. Therefore, $\Braket{M}_{AB}$ intuitively signifies twice the average value of $M$ at the initial and final configurations within their respective canonical ensembles.

Similarly, we introduce a special average at configuration $\vec{q}_{J}$, $\Braket{\quad}_{J_{Z}}$, for any configurational function $W$, expressed as
\begin{eqnarray}
\Braket{W\left( \vec{q} \right)}_{J_{Z}} = \sum_{\vec{q}'} R\left( \vec{q}'|\vec{q}_{J} \right) \cdot W\left( \vec{q}' \right).
\end{eqnarray}

Given that heat inflow from the thermal bath to the system, $Q$, is a fundamental thermodynamic quantity, we apply the special average of $\Braket{\quad}_{AB}$ in Eq.~\eqref{eq:VA} to the bath entropy change, yielding
\begin{widetext}
\begin{eqnarray}
\label{eq:va-dsb}
\Braket{\Delta\tilde{S}_{\textrm{b}}}_{AB} =D_{\textrm{NOL}}^{A} - D_{\textrm{NOL}}^{B} + \sum_{\vec{q}_{A}}R\left( \vec{q}_{A}|\vec{q}_{B} \right) \ln \frac{R\left( \vec{q}_{B}|\vec{q}_{A} \right)}{ R_{\textrm{G}}\left( \vec{q}_{B}|\vec{q}_{A} \right) } - \sum_{\vec{q}_{B}}R\left( \vec{q}_{B}|\vec{q}_{A} \right) \ln \frac{R\left( \vec{q}_{A}|\vec{q}_{B} \right)}{ R_{\textrm{G}}\left( \vec{q}_{A}|\vec{q}_{B} \right) }. 
\end{eqnarray}
\end{widetext}
When we applied the following relationships to the transition probability: 
\begin{eqnarray}
\ln R\left( \vec{q}_{B} | \vec{q}_{A} \right) &=& \ln g\left( \vec{q}_{B} \right) - \beta\left( \vec{q}_{B}\cdot\vec{V}_{A}\left( \vec{q}_{A} \right) \right) + \beta f_{A}\left( \vec{q}_{A} \right) \nonumber \\
\end{eqnarray}
{and the relationships for $\Braket{\quad}_{AB}$ of $f$:
\begin{eqnarray}
\Braket{\beta\tilde{f}_{B}\left( \vec{q}_{B} \right)}_{AB} &=& \Braket{\beta\tilde{f}_{B}\left( \vec{q}_{B} \right)}_{A} + \beta \tilde{f}_{B}\left( \vec{q}_{B} \right) \nonumber \\
\Braket{\beta\tilde{f}_{A}\left( \vec{q}_{A} \right)}_{AB} &=& \Braket{\beta\tilde{f}_{A}\left( \vec{q}_{A} \right)}_{B} + \beta \tilde{f}_{A}\left( \vec{q}_{A} \right)
\end{eqnarray}
to Eq.~\eqref{eq:va-dsb}, we obtain 
\begin{eqnarray}
\Braket{\Delta\tilde{S}_{\textrm{b}}}_{AB} = -\Delta D_{\textrm{NOL}} + \Delta\tilde{d} -\beta\Braket{\Delta\tilde{f}}_{AB} + \beta\Delta\tilde{f}.
\end{eqnarray}
Again, note that $\Delta \tilde{f}$ does not correspond to the variations in the free energy \textit{after} the transformation, and the information on $f$ is excluded. 
To this end, we first employed $\Delta d$ to obtain the following relationship with $\Delta f$
\begin{widetext}
\begin{eqnarray}
\label{eq:sen-F}
\Delta S_{\textrm{b}} + \Delta d = -{\beta} \left( f_{B}  - f_{A}\right) + \beta \left( \vec{q}_{A}\cdot \vec{V}_{B} - \vec{q}_{B}\cdot \vec{V}_{A} \right) = -\beta \Delta f,
\end{eqnarray}
\end{widetext}
where the last equation can be obtained because $\Gamma$ should be symmetrical by definition. 
Using Eq.~\eqref{eq:sen-F}, and the relationship between the
\begin{eqnarray}
-\Delta \tilde{S}_{\textrm{b}} = \beta\tilde{Q}, 
\end{eqnarray}
the following desired relationship can be obtained: 
\begin{eqnarray}
\label{eq:nol-Q}
\Delta D_{\textrm{NOL}} - \left[ 2\Delta \tilde{d} - \Braket{\Delta\tilde{d}}_{AB} \right] = \beta \tilde{Q}.
\end{eqnarray}
We first consider the physical significance of $\Delta \tilde{d}$, which can be rewritten as the variations in stochastic relative entropy: 
\begin{eqnarray}
\Delta \tilde{d} &=& d_{\textrm{KL}}\left[ R\left( \vec{q}_{B}|\vec{q}_{0}\right) : R_{\textrm{G}}\left( \vec{q}_{B}|\vec{q}_{0} \right) \right] \nonumber \\ 
&-& d_{\textrm{KL}}\left[ R\left( \vec{q}_{A}|\vec{q}_{0}\right) : R_{\textrm{G}}\left( \vec{q}_{A}|\vec{q}_{0} \right) \right]. 
\end{eqnarray}
Therefore, $\Delta \tilde{d}$ corresponds to the variance in the nonlinearity contribution to $\vec{q}_{0}$ from between the initial and final configurations. This variance is discernible from the non-local nonlinearity at $\vec{q}_{0}$ as 
\begin{eqnarray}
D_{\textrm{NOL}}^{0} = \sum_{\vec{q}} R\left( \vec{q}|\vec{q}_{0} \right)\cdot d_{\textrm{KL}}\left[ R\left( \vec{q}|\vec{q}_{0} \right) : R_{\textrm{G}}\left( \vec{q}|\vec{q}_{0} \right) \right].
\end{eqnarray}
Based on the preceding discussion and Eq.~\eqref{eq:nol-Q}, we observed that in transforming the system transition driven by local nonlinearity into a system transition in thermal contact, we can ascertain the connections between the changes in nonlinearity, $\Delta D_{\textrm{NOL}}$ (on a statistical manifold), and a thermodynamic variable (in this context, heat inflow $Q$), which 
is initially prompted by the nonlinearity of the configuration space.

We further examined the significance of the term $\left[ 2\Delta \tilde{d} - \Braket{\Delta\tilde{d}}_{AB} \right]$ in Eq.~\eqref{eq:nol-Q} by qualitatively considering the (sufficient) conditions under which the term becomes zero. Based on the definition of $\Braket{\quad}_{AB}$ in Eq.~\eqref{eq:VA}, 
\begin{eqnarray}
\vec{q}_{B} &=& \vec{q}_{A} + \vec{H}\left( \vec{q}_{A} \right) \nonumber \\
\vec{q}_{A} &=& \vec{q}_{B} + \vec{H}\left( \vec{q}_{B} \right)
\end{eqnarray}
are satisfied by the corresponding canonical distributions $c_{A}$ and $c_{B}$ respectively, which exhibit sharp peaks around $\vec{q}_{B}$ and $\vec{q}_{A}$, 
\begin{eqnarray}
\label{eq:va-mean}
\left[ 2M - \Braket{M}_{AB} \right] \simeq 0
\end{eqnarray}
typically holds for any function $M$ in both initial and final configurations. Based on this perspective, 
$\left[ 2\Delta \tilde{d} - \Braket{\Delta\tilde{d}}_{AB} \right]$ intuitively corresponds to the \textit{ noncyclic } contribution through partial nonlinearity at the random configuration.

Here, we note that the proposed thermodynamic treatment can be generalized to any system when (i) the initial and final points of each vector correlate to the average of an appropriate map (in this case, $\phi_{\textrm{th}}$) for a specified probability distribution (here, CDOS), thereby conceptualizing it as a dynamical system, and (ii) we introduce the KL divergence between probability distributions derived through the map for the initial and final points. 

The rationale for applying this treatment to the canonical ensemble is two-fold: (i) there is a need to devise a new method to formulate nonlinearity characteristics across multiple configurations, and (ii) determine what additional information is required for unifying the concept of the vector field and KL divergence. In the current context, this pertains to the nonlinearity information at the random configuration, where $\Delta \tilde{d}$ poses significant influence.

\subsubsection*{Averaged Nonlinearity Character over Multiple Configurations}

Based on the above discussion, we are now ready to address the averaged nonlinearity characteristic over \textit{multiple} configurations, which has been difficult to formulate using existing approaches. 
To achieve this, we first examined how the second law of thermodynamics and the integral fluctuation theorem are expressed for the present transformation for nonlinearity. From the above equations, we obtain 
\begin{eqnarray}
\tilde{\sigma} = \ln \frac{P_{+}\left( \vec{q}_{A},\vec{q}_{B} \right) P_{\textrm{G}-}\left( \vec{q}_{A},\vec{q}_{B} \right)  }{P_{-}\left( \vec{q}_{A},\vec{q}_{B} \right) P_{\textrm{G}+}\left( \vec{q}_{A},\vec{q}_{B} \right) }.
\end{eqnarray}
The ensemble average for negative entropy production along the forward transition $P_{+}$ is given by 
\begin{eqnarray}
\label{eq:sigp}
-\Braket{\tilde{\sigma}}_{P+} &=& \Braket{ \ln \frac{P_{\textrm{G}+}P_{-}}{P_{\textrm{G}-}P_{+} } }_{P_{+}}  \le \ln \Braket{\frac{P_{\textrm{G}+}P_{-}}{P_{\textrm{G}-}P_{+} }}_{P_{+}} \nonumber \\
&=& \ln \sum_{\vec{q}_{A},\vec{q}_{B}} \frac{P_{\textrm{G}+}P_{-}}{P_{\textrm{G}-} } = \ln\Braket{e^{\sigma_{\textrm{G}}}}_{P-},
\end{eqnarray}
where Jensen's inequality was employed. 
Therefore, we obtain the corresponding second law of thermodynamics as follows:
\begin{eqnarray}
\label{eq:2nd}
\Braket{\tilde{\sigma}}_{P+} \ge -\ln\Braket{e^{\sigma_{\textrm{G}}}}_{P-}.
\end{eqnarray}
Similarly, considering the ensemble average of the exponential of the entropy production, we obtain the integral fluctuation theorem as
\begin{eqnarray}
\label{eq:if}
\Braket{e^{-\tilde{\sigma}}}_{P+} = \Braket{e^{\sigma_{\textrm{G}}}}_{P-}.
\end{eqnarray}
From the above equations, we see that the backward transition average of entropy production for a linear system, $\Braket{e^{\sigma_{G}}}_{P-}$ acts as the identity of $\mathbf{1}$; that is, under this extended interpretation, Eqs.~\eqref{eq:2nd} and~\eqref{eq:if} correspond to the typical second law and integral fluctuation theorem of $\Braket{\tilde{\sigma}}_{P+} \ge 0$ and $\Braket{e^{-\tilde{\sigma}}}_{P+} = 1$.

\begin{figure}[h]
\begin{center}
\includegraphics[width=0.6\linewidth]{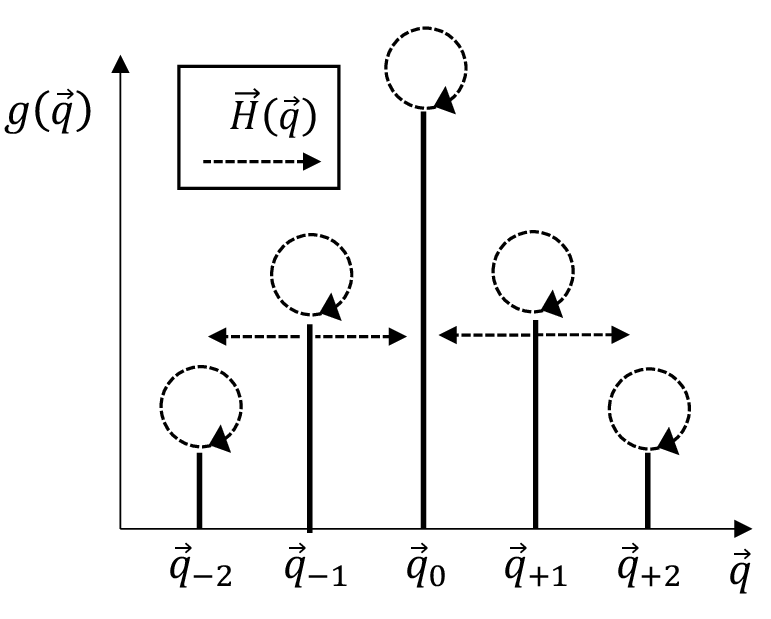}
\caption{Schematic illustration of coarse-grained configuration space with the resultant CDOS $g\left(q\right)$ (vertical bold lines) and vector field $\vec{H}$ (dashed curves) for equiatomic binary alloys on representative lattice (including fcc and bcc). }
\label{fig:trans-area}
\end{center}
\end{figure}

Based on the above discussion, we can examine the average nonlinearity change $\Delta D_{\textrm{NOL}}$ over multiple configurations using the introduced thermodynamic quantities as follows:
First, Eq.~\eqref{eq:nol-Q} can be rewritten as follows:
\begin{eqnarray}
\label{eq:dnol}
\Delta D_{\textrm{NOL}} - \Delta \tilde{d} + \Braket{\Delta\tilde{d}}_{AB} = -\left( \Delta\tilde{S}_{\textrm{b}} - \Delta\tilde{d} \right).
\end{eqnarray}
By applying the average $\Braket{\quad}_{P^{0}_{+}}$ to the right side of Eq.~\eqref{eq:dnol} and using Eq.~\eqref{eq:p0}, we immediately obtain the inequality (in a manner similar to the above derivation for the second law).
\begin{eqnarray}
-\Braket{\Delta\tilde{S}_{\textrm{b}} - \Delta\tilde{d} }_{P_{+}^{0}} \le \ln\Braket{\frac{P^{0}_{\textrm{G}-}}{P^{0}_{\textrm{G}+} }}_{P^{0}_{+}} = \ln\Braket{e^{-\sigma^{0}_{\textrm{G}}}}_{P_{+}^{0}}.
\end{eqnarray}
Applying the above inequality and relationships between $\Braket{\quad}_{AB}$, $\Braket{\quad}_{A}$ and $\Braket{\quad}_{B}$ in Eq.~\eqref{eq:VA} to Eq.~\eqref{eq:dnol}, we obtain the upper bound for the average nonlinearity change as
\begin{eqnarray}
\label{eq:nol-change}
\Braket{\Delta \check{D}_{\textrm{NOL}}}_{P_{+}^{0}} \le \ln\Braket{e^{-\sigma_{\textrm{G}}^{0}}}_{P_{+}^{0}},
\end{eqnarray}
where
\begin{eqnarray}
\check{D}_{\textrm{NOL}}^{K} &=& D_{\textrm{NOL}}^{K} - \Braket{\tilde{d}_{I}}_{K_{Z}}.
\end{eqnarray}
Equality condition for Eq.~\eqref{eq:nol-change} is given by
\begin{eqnarray}
^{\forall}\left(\vec{q}_{A},\vec{q}_{B}\right): \, \dfrac{P^{0}_{\textrm{G}+}\left(\vec{q}_{A},\vec{q}_{B}\right)P^{0}_{-}\left(\vec{q}_{A},\vec{q}_{B}\right)}{P^{0}_{\textrm{G}-}\left(\vec{q}_{A},\vec{q}_{B}\right)P^{0}_{+}\left(\vec{q}_{A},\vec{q}_{B}\right)} = c \,
\end{eqnarray}
where $c$ is a real constant depending only on the practical CDOS. This directly means that the upper bound is achieved when entropy production for practical and linear system retains constant difference for any transition.
\begin{figure}[h]
\begin{center}
\includegraphics[width=0.95\linewidth]{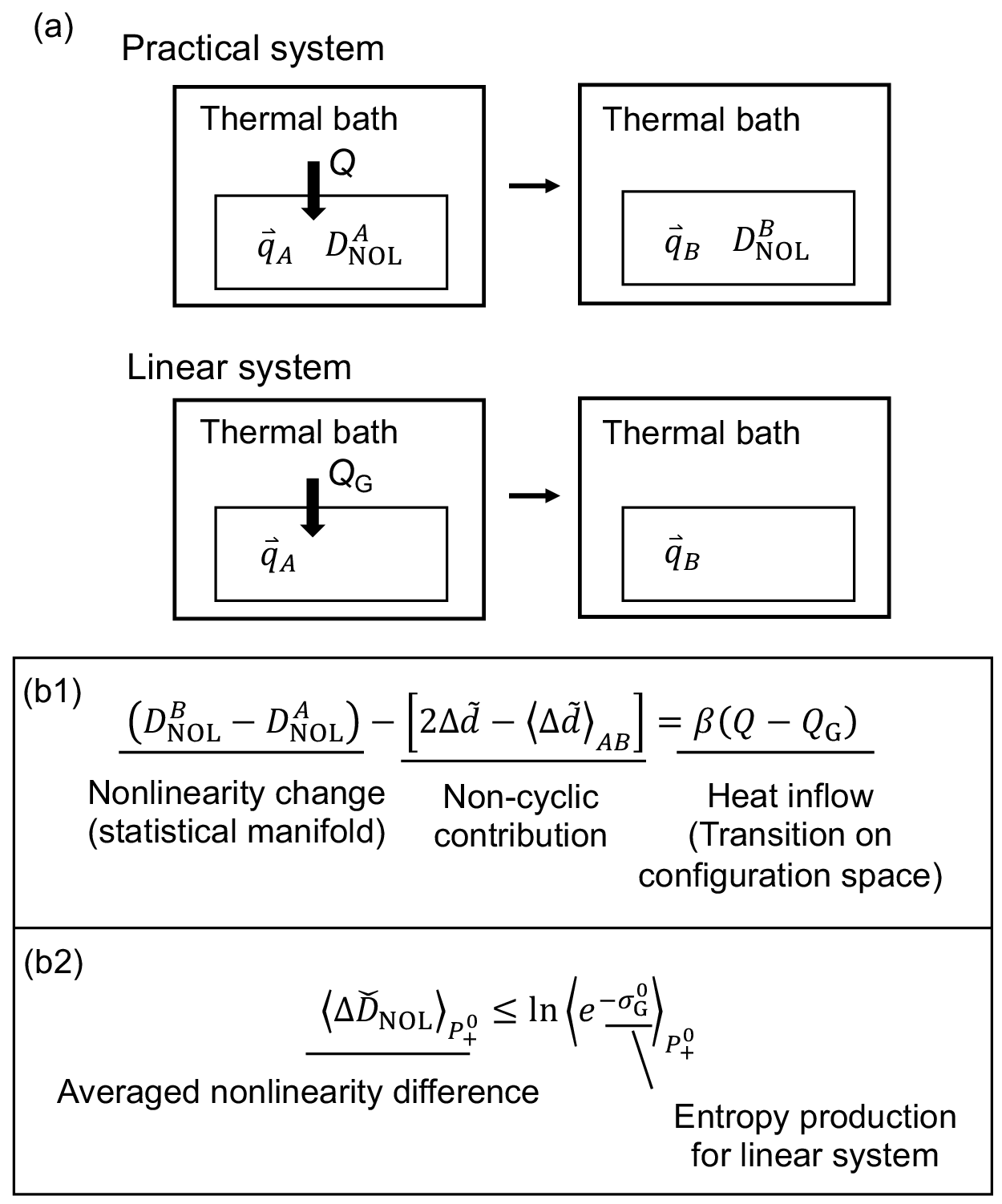}
\caption{Summary of the representative results for nonlinearity character obtained through the present transformation. (a) Brief concept of the present setup, measuring the transformed thermodynamic quantities on practical system from those on linear system. The practical and linear system exhibit transition from initial $\vec{q}_{A}$ to final configuration $\vec{q}_{B}$, respectively, contacting with a thermal bath at inverse temperature $\beta$. (b1) Relationship between $\Delta D_{\textrm{NOL}} = D_{\textrm{NOL}}^{B}-D_{\textrm{NOL}}^{A}$ and heat inflow to the system, bridging the concept of nonlinearity on statistical manifold and nonlinearity on configuration space through the transformed thermodynamic quantities (corresponding to Eq.~\eqref{eq:nol-Q}).  (b2) From (b1), upper bound for average nonlinearity along forward transition is clarified, which is characterized by entropy production for the linear system (corresponding to Eq.~\eqref{eq:nol-change}).  }
\label{fig:thermo-nol}
\end{center}
\end{figure}

Currently, the exact interpretation of Eq.~\eqref{eq:nol-change} in terms of the lattice geometry is challenging, and thus, we can provide an intuition for Eq.~\eqref{eq:nol-change} based on the following features of the vector field typically holds true for substitutional alloys.
We first consider the term $\check{D}_{\textrm{NOL}}^{K}$, which is expressed explicitly as follows:
\begin{eqnarray}
\check{D}^{K}_{\textrm{NOL}} &=& \sum_{\vec{q}_{I}} R\left( \vec{q}_{I} | \vec{q}_{K} \right) \ln\frac{R\left( \vec{q}_{I} | \vec{q}_{K} \right) }{ R_{\textrm{G}}\left( \vec{q}_{I} | \vec{q}_{K} \right)} \nonumber \\
&-& \sum_{\vec{q}_{I}} R\left( \vec{q}_{I} | \vec{q}_{K} \right) \ln\frac{R\left( \vec{q}_{I} | \vec{q}_{0} \right) }{ R_{\textrm{G}}\left( \vec{q}_{I} | \vec{q}_{0} \right)}.
\end{eqnarray}
Therefore, $\check{D}^{K}_{\textrm{NOL}}$ corresponds to non-local nonlinearity at configuration $\vec{q}_{k}$, $D_{\textrm{NOL}}^{K}$, measured from partial nonlinearity contribution from random configuration, $\ln\left[R\left( \vec{q}_{I} | \vec{q}_{0} \right) / R_{\textrm{G}}\left( \vec{q}_{I} | \vec{q}_{0} \right)  \right]$, averaged over canonical distribution with many-body interaction at configuration $\vec{q}_{k}$, $\vec{V}_{k}$, as expressed in Eq.~\eqref{eq:gp}. 
We proceed to intuitively interpret the average $\Braket{\Delta \check{D}_{\textrm{NOL}}}{P_{+}^{0}}$ through a simplified model: In Fig.~\ref{fig:trans-area}, we present schematic representations of the coarse-grained configuration space along with the corresponding CDOS (depicted as vertical bold lines) and their vector field $\vec{H}\left( \vec{q} \right)$ (illustrated by dashed arrows). These representations are applicable to equiatomic binary systems on typical lattices, such as fcc and bcc across various pair correlations.\cite{fcc-geom}

Here, $\vec{q}_{0}$ represents near-random configurations, $\vec{q}_{\pm 2}$ denotes around ground-state ordered configurations, and $\vec{q}_{\pm 1}$ indicates partially ordered configurations. This coarse-graining approach is further validated through dynamic mode decomposition of vector field alterations for multiple systems with varying coordination numbers of pair correlations. As observed, the dominant mode for nonlinearity exhibits distinct changes in three regions: $\vec{q}_{0}$, $\vec{q}_{\pm 1}$, and $\vec{q}_{\pm 2}$.\cite{cls}

The common features of the vector field, as identified from these studies, are: (i) around $\vec{q}_{0}$ and $\vec{q}_{\pm 2}$, $\vec{H}$ approximates to zero, primarily because of the locally linear nature of $\phi_{\rm{th}}$ near $\vec{q}_{0}$ and its role as an adsorption point near $\vec{q}_{\pm 2}$, supported by the fact that multiple candidates of many-body interactions provide the ordered configuration at thermodynamic equilibrium,\cite{CPG} and (ii) $\vec{H}$ around $\vec{q}_{\pm 1}$ has endpoints near $\vec{q}_{\pm 2}$, $\vec{q}_{\pm 1}$ or $\vec{q}_{0}$ (double sign correspondence). Consequently, we set the transition probabilities of
\begin{eqnarray}
\label{eq:tps}
R\left(\vec{q}_{0}|\vec{q}_{0}\right) &=& R\left(\vec{q}_{\pm 2}|\vec{q}_{\pm 2}\right) = 1 \nonumber \\
R\left(\vec{q}_{K}|\vec{q}_{+1}\right) &=& R_{K+1} > 0 \nonumber \\
R\left(\vec{q}_{K}|\vec{q}_{-1}\right) &=& R_{K-1} > 0 \nonumber \\
R\left(\vec{q}_{i}|\vec{q}_{k}\right) &=& 0 \quad \left(\textrm{otherwise}\right).
\end{eqnarray}
Applying Eq.~\eqref{eq:tps} to the average of $\Braket{\Delta \check{D}_{\textrm{NOL}}}_{P_{+}^{0}}$, we obtain
\begin{eqnarray}
\label{eq:ddnol-ave}
\Braket{\Delta \check{D}_{\textrm{NOL}}}_{P_{+}^{0}} &=& \Braket{\check{D}_{\textrm{NOL}}^{B} - \check{D}_{\textrm{NOL}}^{A}}_{P_{+}^{0}} \nonumber \\
&=& \sum_{K=0,+2}\left[ \check{D}_{\textrm{NOL}}^{K} - \check{D}_{\textrm{NOL}}^{+1} \right] R_{K+1}g\left(\vec{q}_{+1}\right) \nonumber \\
&+& \sum_{K=0,-2}\left[ \check{D}_{\textrm{NOL}}^{K} - \check{D}_{\textrm{NOL}}^{-1} \right] R_{K-1}g\left(\vec{q}_{-1}\right). 
\end{eqnarray}
Eq.~\eqref{eq:ddnol-ave} certainly indicates that $\Braket{\Delta \check{D}_{\textrm{NOL}}}_{P_{+}^{0}}$ corresponds to the difference in averaged nonlinearity between partially ordered and other configurations. The coefficient $R_{K\pm1}g\left( \vec{q}_{\pm1}\right)$ naturally provides the weight for the nonlinearity disparity through density of states in the configuration space, including the effects of stochastic transitions: When the transition is substantially deterministic, the weight merely becomes CDOS itself, and when the transition is stochastic, the weight further includes the effects of corresponding fluctuation of the nonlinearity. 

Regarding the right-hand side of Eq.~\eqref{eq:nol-change}, since $\sigma_{\textrm{G}}^{0}$ pertains to entropy production in linear systems, it is completely determined by the covariance matrix of the CDOS for a practical system. We emphasize here that r.h.s. of Eq.~\eqref{eq:nol-change} includes entropy production for linear system, while its forward average $P_{+}^{0}$ is taken along practical transition. Since from the above discussion, the linear system merely requires CDOS but not practical lattice itself, estimation of the r.h.s. of Eq.~\eqref{eq:nol-change} is straightforward. For a concrete example, we employ the simplified configuration space of Fig.~\ref{fig:trans-area} and corresponding transition probabilities of Eq.~\eqref{eq:tps}. Under these conditions, we obtain
\begin{eqnarray}
\Braket{e^{-\sigma_{\textrm{G}}^{0}}}_{P_{+}^{0}} &=& \sum_{K=0,+2}\left[ e^{-\sigma_{\textrm{G}}^{0}\left( \vec{q}_{+1}, \vec{q}_{K} \right)}    \right] R_{K+1}g\left(\vec{q}_{+1}\right) \nonumber \\
&+& \sum_{K=0,-2}\left[ e^{-\sigma_{\textrm{G}}^{0}\left( \vec{q}_{-1}, \vec{q}_{K} \right)}   \right] R_{K-1}g\left(\vec{q}_{-1}\right),
\end{eqnarray}
where
\begin{eqnarray}
\label{eq:pg}
\sigma_{\textrm{G}}^{0}\left( \vec{q}_{K'},  \vec{q}_{K}\right) = d_{\textrm{KL}}\left( P_{\textrm{G+}}^{0}, P_{\textrm{G-}}^{0} \right).
\end{eqnarray}
It is clear that r.h.s. of Eq.~\eqref{eq:pg} can be fully determined by the information about landscape of Gaussian CDOS for the linear system, with probabilistic average for transition of practical system. 

According to this discourse, Eq.~\eqref{eq:nol-change} suggests that the difference in averaged nonlinearity between partially ordered and other configurations, measured from the nonlinearity information at a random configuration, is constrained by details about the covariance matrix of CDOS for a practical system, where this matrix can be deduced from the number of multisite figures in the specified lattice for the corresponding correlation.\cite{mom}

In summary, Figure ~\ref{fig:thermo-nol} demonstrates the representative outcomes for nonlinearity in the canonical ensemble derived in this study. These include (i) Eq.~\eqref{eq:nol-Q}, which integrates the concept of nonlinearity as a vector field $\vec{H}$ and KL divergence $D_{\textrm{NOL}}$ (as depicted in Fig.~\ref{fig:thermo-nol} (b1)), and (ii) Eq.~\eqref{eq:nol-change}, indicating that the variance in average nonlinearity between partially ordered and other configurations is bound by entropy production (i.e., the covariance matrix of CDOS).

These findings strongly suggest that to further explore nonlinearity in the canonical ensemble for \textit{multiple} configurations, the proposed thermodynamic approach holds considerable approach.

\section{Conclusions}
This study proposed a thermodynamic approach to address nonlinearity within a canonical ensemble, framing the system transition driven by nonlinearity as a transition in thermal contact, guided by principles of stochastic thermodynamics. This transformative process reveals that variations in nonlinearity, quantified as KL divergence on a statistical manifold and originating from the non-cyclic contribution of partial nonlinearity at a perfectly random configuration, align with the equivalent thermodynamic variable of heat inflow, initially prompted by the nonlinearity of the configuration space. We clarify that the average alteration in nonlinearity on the statistical manifold is restricted by the entropy production in a linear system. This study paves the avenue for advancing understanding of the nonlinearity in canonical ensembles through the application of the proposed thermodynamic framework.

\section{Acknowledgement}
The author expresses cordial thanks to Yoshikazu Tabata at Kyoto University for fruitful discussions. 
This work was supported by Grant-in-Aids for Scientific Research on Innovative Areas on High Entropy Alloys through the grant number JP18H05453 and a Grant-in-Aid for Scientific Research (23K04359) from the MEXT of Japan, and Research Grant from Hitachi Metals$\cdot$Materials Science Foundation.

\end{document}